\documentclass[%
 reprint,
 amsmath,amssymb,
 aps,
]{revtex4-2}

\usepackage{graphicx}
\usepackage{dcolumn}
\usepackage{bm}

\begin{document}

\preprint{APS/123-QED}

\title{Cavity locking with spatial modulation of optical phase front for laser stabilization}
\thanks{A footnote to the article title}%

\author{Sheng Feng}
 \email{fensf2a@hust.edu.cn}
\author{Songqing You}%
\altaffiliation[Also at ]{Hubei Key Laboratory of Modern Manufacturing Quantity Engineering, School of Mechanical Engineering, Hubei University of Technology, Wuhan, Hubei 430068, China.}
\author{Peng Yang}
\author{Fenglei Zhang}
\author{Yunlong Sun}
\affiliation{%
 School of Electrical and Electronic Infomation Engineering, Hubei Polytechnic University, Huangshi, Hubei 435003, P.R. China
}%

\author{Boya Xie}
\affiliation{Hubei Key Laboratory of Modern Manufacturing Quantity Engineering, School of Mechanical Engineering, Hubei University of Technology, Wuhan, Hubei 430068, China.}

\date{\today}

\begin{abstract}
We study optical cavity locking for laser stabilization through spatial modulation of the phase front of a light beam. A theoretical description of the underlying principle is developed for this method and special attention is paid to residual amplitude modulation (RAM) caused by experimental imperfections, especially the manufacture errors of the spatial phase modulator. The studied locking method owns the common advantages of the Pound-Drever-Hall method and the tilt-locking one, and it can provide a more artful way to eliminate RAM noise in phase modulation for the ultimate stability of lasers. In situations where cost and portability are a practical issue, the studied method allows one to realize compact laser stabilization systems locked to  Fabry-P$\acute{\mbox{e}}$rot cavities without use of expensive bulky devices, such as signal generators and electro-optic modulators.
\end{abstract}

\maketitle


Ultra-stable lasers with high-spectral purity play an essential role in a variety of advanced research fields, such as optical atomic clocks \cite{Hinkley2013,Bloom2014,ludlow2015}, tests of relativity
\cite{eisele2009,Wiens2016}, gravitational wave detectors \cite{ligo2009}, and photonic microwave synthesizers \cite{Fortier2011}. State-of-the-art stable lasers locked to high-finesse Fabry P$\acute{\mbox{e}}$rot (FP)
cavities delivering single-mode electromagnetic waves have been realized with integral linewidths below 1 Hz \cite{Matei2017,Robinson2019}. Another research direction of relevance is to construct portable stable laser systems
\cite{Davila2017} towards applications wherein robustness and integration are primary concerns, with a 20-ms stability of $10^{-13}$ already reported \cite{Zhang2020}. Despite several locking schemes of potential interest
\cite{hansch1980,Shaddock2000,Thorpe2008,yang2017,wang2019}, most of these performance demonstrations have been accomplished by use of the standard Pound-Drever-Hall (PDH) one \cite{drever1983} in which the phase front of the
optical carrier undergoes temporal modulation that generates radio frequency (r.f.) sidebands.

A well-known issue with the PDH modulation technique is residual amplitude modulation (RAM) that arises when the r.f. modulation sidebands are unequal in magnitude, not exactly out of phase, or both. RAM couples frequency
offset noise to the servo error signal and thereby aggravates the laser frequency instability. A few RAM reduction schemes \cite{Wong1985,Muller2003,Burck2003,LLi2012} have been implemented with the most remarkable one
demonstrated recently \cite{Zhang2014} that utilized an active servo loop involving both DC electric field and temperature corrections applied to the phase modulator. RAM control on the $10^{-6}$ level was reported with a
RAM-induced frequency instability comparable to or lower than thermal noise limit \cite{Zhang2014}.

Instead of using {\it temporal} modulation as the PDH method does \cite{drever1983}, we explore in this work {\it spatial} modulation of the optical phase front of the carrier for FP cavity locking. Aside from its own merit of
fundamental interest, the studied cavity locking method provides handy ways for RAM reduction; moreover, it has the common advantages of the PDH method and the tilt-locking one the later of which nonetheless utilizes
travelling-wave cavities for laser stabilization \cite{Chabbra2021}. In comparison with the PDH one, the studied method can provide larger error signals for the servo loop and may eliminate the need of electro-optic modulators
(EOM's) and signal generators that are expensive, bulky, and otherwise necessary for temporal phase modulation, which is beneficiary to low-cost portable laser stabilization systems locked to FP cavities
\cite{Davila2017,Zhang2020}.

\begin{figure}[htp]
\centering
\includegraphics[width=6.0cm]{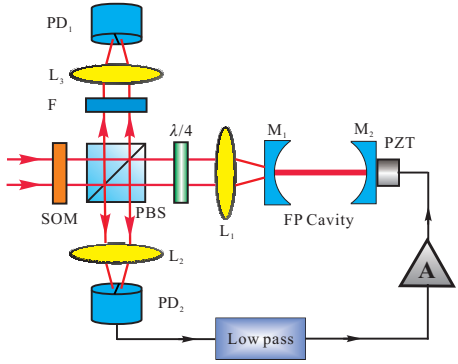}
\caption{(color online) Schematics for cavity locking by spatial modulation of optical phase front. SOM: Spatio-optic modulator. PBS: Polarizing beam-splitter. $\lambda/4$: Quarter wave plate. L$_1$: Focal lens coupling the
incident light into the cavity. $M_{1,2}$: Cavity mirrors. PZT:  Piezo-electrical transducer. L$_{2,3}$: Imaging lenses used to map the SOM images to the corresponding photo-detectors. PD$_{1,2}$: Dual-quadrant
photo-detectors. F: An optional optical element used to tailor the spatial mode amplitudes of the light field received by PD$_{1}$ for RAM cancellation.} \label{fig:model}
\end{figure}

In what follows, let describe the principle underlying the studied cavity locking method through spatial modulation of optical phase front. For simplicity without loss of generality, let consider a collimated laser beam in a
Gaussian mode, $E_{\mbox{i}}(\vec{r},t)$, that is coupled into an FP cavity through a focal lens (Fig. \ref{fig:model}). A transparent optical element, namely a spatio-optic modulator (SOM), is inserted into the incident light
beam before the lens such that the phase front of the beam undergoes spatial modulation. The spatially modulated light field $E_{\mbox{m}}(\vec{r},t)$ is described as,
\begin{equation} \label{eq:modulation}
E_{\mbox{m}}(\vec{r},t)=E_{\mbox{i}}(\vec{r},t)U(\vec{r})=E_{\mbox{i}}(\vec{r},t)e^{i\varphi(\vec{r})} \ ,
\end{equation}
wherein $U(\vec{r})$ is the SOM modulation function, and the vector $\vec{r}$ represents the two-dimensional  coordinates in the plane perpendicular to the propagation axis of the light beam. Unity transmittance is assumed for
the SOM and $\varphi(\vec{r})$ is the spatial distribution of the change to the light field phase front caused by the SOM.

To get an insight into the cavity locking method, let assume $|\varphi(\vec{r})|<<1$ for the moment. Under this assumption, the modulated light field $E_{\mbox{m}}(\vec{r},t)$ reads
\begin{eqnarray} \label{eq:mod}
E_{\mbox{m}}(\vec{r},t)&=&E_{\mbox{i}}(\vec{r},t)+i\varphi(\vec{r})E_{\mbox{i}}(\vec{r},t) \ ,
\end{eqnarray}
from which it follows that the modulated light field becomes a superposition of the incident light field, $E_{\mbox{i}}(\vec{r},t)$, and another effective field, $E_{\mbox{e}}(\vec{r},t)\equiv
i\varphi(\vec{r})E_{\mbox{i}}(\vec{r},t)$, as a consequence of spatial modulation. For the purpose of this work, the SOM may be designed for $\varphi(\vec{r})$ to satisfy
\begin{equation} \label{eq:elementdesign}
\iint d\vec{r} E_{\mbox{i}}^*(\vec{r},t) E_{\mbox{e}}(\vec{r},t) = \iint d\vec{r} \varphi(\vec{r}) |E_{\mbox{i}}(\vec{r},t)|^2 = 0 \ ,
\end{equation}
in which the integral area covers the whole cross section of the incident beam. In other words, the spatial mode of the effective field, $E_{\mbox{e}}(\vec{r},t)$, is orthogonal to that of the incident field,
$E_{\mbox{i}}(\vec{r},t)$, and hence the two fields may not simultaneously resonate inside a narrow-linewidth optical cavity due to their spatial mode orthogonality \cite{Siegman1986}.

Now let suppose that the incident field $E_{\mbox{i}}(\vec{r},t)$ is near resonance in the FP cavity of Fig. \ref{fig:model}, and that the cavity linewidth is narrow enough so that the effective field $E_{\mbox{e}}(\vec{r},t)$
is far-off resonance. Then after reflection from the cavity, the modulated light field becomes \cite{hansch1980}
\begin{eqnarray} \label{eq:reflection}
E_{\mbox{r}}(\vec{r},t)&=&\left\{\sqrt{R_1}\left[1+i\varphi(\vec{r})\right] -\frac{T_1}{\sqrt{R_1}}\frac{e^{i\delta}\sqrt{R}}{1-e^{i\delta}\sqrt{R}} \right\}E_{\mbox{i}}(\vec{r},t) \nonumber \\
&=& \left[\chi e^{i\theta} + i\varphi(\vec{r})\right] \sqrt{R_1}\ E_{\mbox{i}}(\vec{r},t) \ ,
\end{eqnarray}
in which
\begin{equation}
\chi e^{i\theta}\equiv 1- \frac{T_1}{R_1}\frac{e^{i\delta}\sqrt{R}}{1-e^{i\delta}\sqrt{R}} \ ,
\end{equation}
and $R_1, T_1$ are respectively the reflectivity and transmissivity of the cavity entrance mirror M$_1$, $R\equiv R_1 R_2$ ($R_2$ the reflectivity of the other cavity mirror M$_2$), and $\delta$ stands for the round trip phase
delay of the light field inside the cavity.

\begin{figure}[htp]
\centering
\includegraphics[width=3.5cm]{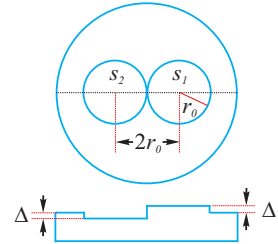}
\caption{(color online) An example of a designed SOM for cavity locking. $\Delta\equiv (\varphi_0/2\pi)\cdot \lambda/(n-1)$ with $\lambda$ being the wavelength of the light beam and $n$ the refractive index of the SOM
material). (Top): Front view of the SOM. (Bottom): The SOM surface profile along the dash line as indicated on the top.} \label{fig:som}
\end{figure}

In what follows, we will show how to exploit an optical element of spatial modulation, i.e. an SOM, to implement cavity locking based on Eq. (\ref{eq:reflection}). Given that the incident light field $E_{\mbox{i}}(\vec{r},t)$
is in a Gaussian mode which is symmetric with respect to the light propagation axis,
\begin{equation}\label{eq:fieldsym}
E_{\mbox{i}}(\vec{r},t)= E_{\mbox{i}}(-\vec{r},t)\ ,
\end{equation}
one may utilize an SOM with a spatial modulation function that introduces an  anti-symmetric phase change to the incident beam around its axis,
\begin{equation}\label{eq:phisym}
\varphi(\vec{r})= -\varphi(-\vec{r})\ .
\end{equation}
From Eqs. (\ref{eq:fieldsym}) and (\ref{eq:phisym}) it immediately follows that Eq. (\ref{eq:elementdesign}) holds true and, hence, so does Eq. (\ref{eq:reflection}). An example of a designed SOM satisfying Eq.
(\ref{eq:phisym}) is illustrated in Fig. \ref{fig:som} and the phase front change $\varphi(\vec{r})$ of the light field due to the spatial modulation reads
\begin{equation}\label{eq:phiexample}
\varphi(\vec{r}) = \left\{ \begin{array}{ll}
\pm\varphi_0 & |\vec{r}\mp r_0|<r_0\\
0 & \textrm{otherwise}
\end{array} \right.\ ,
\end{equation}
in which $\pm \varphi_0$ are constant modulation-induced phase changes in the corresponding areas, $S_{1,2}$, which are circular here centered at the points of ($\pm r_0, 0$) respectively with the same radius of $r_0$.


Next we will show how an error signal may be obtained in the cavity locking method. To this end, one may use a dual-quadrant photo-detector to receive the light beam reflected from the cavity. An optical imaging system is
utilized to project the image of the SOM onto the photo-detector, with each quadrant aligned with respect to each of the circles, $S_{1,2}$, on the SOM. The output photo-electric currents from the two quadrants are subtracted
to produce an error signal $\varepsilon(\delta)$ as described by,
\begin{equation}\label{eq:errorsignal}
\varepsilon(\delta) = \iint_{S_1} d\vec{r} |E_{\mbox{r}}(\vec{r},t)|^2 -\iint_{S_2} d\vec{r} |E_{\mbox{r}}(\vec{r},t)|^2\ .
\end{equation}
To make it clearer, one may plug Eqs. (\ref{eq:reflection}) and (\ref{eq:phiexample}) into Eq. (\ref{eq:errorsignal}), leading to
\begin{eqnarray}\label{eq:es}
\varepsilon(\delta) &=& \iint_{S_1} d\vec{r} R_1 |(\chi e^{i\theta}+i\varphi_0)E_{\mbox{i}}(\vec{r},t)|^2 \nonumber \\
&& -\iint_{S_2} d\vec{r} R_1|(\chi e^{i\theta}-i\varphi_0)
E_{\mbox{i}}(\vec{r},t)|^2 \nonumber \\
&=& 4R_1I_0\ \varphi_0 \ \chi\sin\theta  \nonumber \\
&=& 4R_1I_0\ \varphi_0 \ \mbox{Im}\left(1- \frac{T_1}{R_1}\frac{e^{i\delta}\sqrt{R}}{1-e^{i\delta}\sqrt{R}}\right)\ ,
\end{eqnarray}
in which Im($\cdot$) stands for the imaginary part of a complex number and one has invoked Eq. (\ref{eq:fieldsym}) from which it follows that $\iint_{S_1} d\vec{r} |E_{\mbox{i}}(\vec{r},t)|^2=\iint_{S_2} d\vec{r}
|E_{\mbox{i}}(\vec{r},t)|^2\equiv I_0$. The error signal for cavity locking varies as a function of cavity detuning $\delta$ from its peak resonance according to Eq. (\ref{eq:es}).

In what follows, let turn to the RAM problem associated with the studied cavity locking method due to SOM manufacture errors. Specifically, the anti-symmetry given by Eq. (\ref{eq:phisym}) may be broken to some extent in
practice resulting from the difference between the areas, optical transmissivities, modulation depths of $S_1$ and $S_2$; the SOM symmetry breaking, together with unexpected input beam drifting, unbalanced quantum efficiencies
and unequal gains of the two quadrants of the photo-detector, will invalidate the assumptions of both Eq. (\ref{eq:phiexample}) and $\iint_{S_1} d\vec{r} |E_{\mbox{i}}(\vec{r},t)|^2=\iint_{S_2} d\vec{r}
|E_{\mbox{i}}(\vec{r},t)|^2$, causing RAM noises in the error signal. One should note that both the real RAM noises and RAM-like noises are referred to as ``RAM noises" in this work when they cannot be distinguished in
experiment.

To account for these effects in theory, one may define different phase modulations $\varphi_{1,2}$ for each of the two circular areas $S_{1,2}$ respectively with radii $r_{1,2}<r_0$,
\begin{equation}\label{eq:phiasymm}
\varphi(\vec{r}) = \left\{ \begin{array}{ll}
\varphi_1 & |\vec{r} - r_0|<r_1\\
0 & \textrm{otherwise}\\
\varphi_2 & |\vec{r} + r_0|<r_2
\end{array} \right. \ ,
\end{equation}
and photo-currents $I_{1,2}$,
\begin{equation}\label{eq:intensity}
I_1\equiv G_1\iint_{S_1} d\vec{r} |E_{\mbox{i}}(\vec{r},t)|^2,\ \ I_2\equiv G_2\iint_{S_2} d\vec{r} |E_{\mbox{i}}(\vec{r},t)|^2 \ .
\end{equation}
Here the normalized coefficients $G_{1,2}$ account for all the effects resulting from the differences in the optical transmissivities of the areas $S_{1,2}$, beam drifting, the gains and quantum efficiencies of the two
detector quadrants.

With Eqs. (\ref{eq:reflection}), (\ref{eq:phiasymm}), and (\ref{eq:intensity}), the error signal becomes
\begin{eqnarray}\label{eq:esn}
\varepsilon(\delta) &=& G_1\iint_{S_1} d\vec{r} R_1 |(\chi e^{i\theta}+i\varphi_1)E_{\mbox{i}}(\vec{r},t)|^2 \nonumber \\
&& -G_2\iint_{S_2} d\vec{r} R_1|(\chi e^{i\theta}-i\varphi_2)
E_{\mbox{i}}(\vec{r},t)|^2 \nonumber \\
&=& 4R_1(\bar{I}\ \bar{\varphi}+\Delta I\Delta \varphi) \mbox{Im}\left(1- \frac{T_1}{R_1}\frac{e^{i\delta}\sqrt{R}}{1-e^{i\delta}\sqrt{R}}\right) \nonumber \\ && + 2R_1 \left[2\bar{I}\ \bar{\varphi}\Delta \varphi+\Delta
I(\chi^2+\bar{\varphi}^2+(\Delta \varphi)^2)\right]\ ,
\end{eqnarray}
in which $\bar{I}=(I_1+I_2)/2$, $\Delta I=(I_1-I_2)/2$, $\bar{\varphi}=(\varphi_1+\varphi_2)/2$, and $\Delta \varphi=(\varphi_1-\varphi_2)/2$. The first term on the right hand side of Eq. (\ref{eq:esn}) is the error signal
required for cavity locking, whereas the second term describes RAM in the studied scheme that introduces frequency offset noise into the servo error signal leading to degradation of the laser frequency stability.

To get rid of RAM noises in the scheme, one may adjust the coefficients $G_{1,2}$ to change the value of $\Delta I= (I_1-I_2)/2$ according to Eq. (\ref{eq:intensity}) such that
\begin{equation}\label{eq:zeroASOM}
\Delta I=-\frac{2\bar{I}\bar{\varphi}\Delta \varphi}{\chi^2+\bar{\varphi}^2+(\Delta \varphi)^2} \ ,
\end{equation}
which guarantees zero RAM, i.e., the second term in Eq. (\ref{eq:esn}) becomes null. Equality (\ref{eq:zeroASOM}) can be realized without much difficulty in practice by varying the optical transmissivities of the areas
$S_{1,2}$ on the SOM and/or the gains of the two quadrants of the detector. Moreover, unexpected displacement or tilting of the phase-modulated beam can be monitored with another dual-quadrant photo-detector (PD$_1$ in Fig.
\ref{fig:model}) and corrections may be applied to the input beam or the error signal accordingly for further RAM reduction. As for the PDH scheme, to suppress RAM is not a trivial work \cite{Zhang2014}; therefore, our
proposed scheme should be a promising alternative to the PDH one in applications where RAM noise suppression to lower levels may be demanded \cite{Davila2017,Zhang2020}.

\begin{figure}[htp]
\centering
\includegraphics[width=7.0cm]{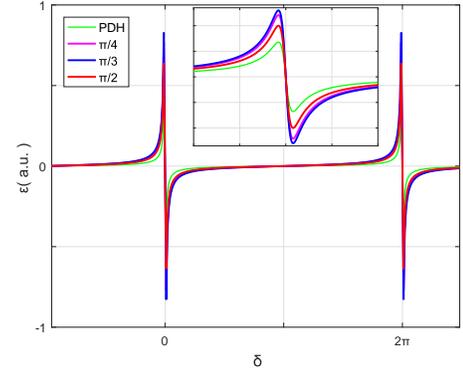}
\caption{(color online) Error signals for cavity locking as a function of the cavity detuning $\delta$ in comparison with that of the PDH scheme. From the simulation results, it follows that the error signal is near its
maximum when $\varphi_0=\pi/3$ (blue curve) and its size is 1.9 times that of the PDH error signal for the modulation depth of 1.08 (green curve). In the simulation, $R_1=99.999\%$ and $R_2=1$ are assumed. Moreover, the total
area of $S_1$ and $S_2$ on the SOM covers half of the cross section of the incident beam.} \label{fig:esstrong}
\end{figure}

In the preceding analysis, it was assumed that $|\varphi(\vec{r})|<<1$, which is nonetheless usually invalid in practice when large error signal sizes are desired. In the following, let relax this assumption and discuss how to
implement cavity locking by spatial modulation of optical phase front. To that end, Eq. (\ref{eq:mod}) may be replaced by
\begin{equation} \label{eq:modstrong}
E_{\mbox{m}}(\vec{r},t)=A E_{\mbox{i}}(\vec{r},t) + E_{\mbox{e}}(\vec{r},t)\ ,
\end{equation}
wherein the effective field now becomes
\begin{equation} \label{eq:efield}
E_{\mbox{e}}(\vec{r},t)= \left[e^{i\varphi(\vec{r})}-A\right]E_{\mbox{i}}(\vec{r},t) \ .
\end{equation}
Here the constant coefficient $A$ is defined as
\begin{equation} \label{eq:constA}
A\equiv \frac{\iint d\vec{r} \cos\varphi(\vec{r})|E_{\mbox{i}}(\vec{r},t)|^2} {\iint d\vec{r} |E_{\mbox{i}}(\vec{r},t)|^2}  \ ,
\end{equation}
which ensures the orthogonality between the effective field  $E_{\mbox{e}}(\vec{r},t)$ and the incident field $E_{\mbox{i}}(\vec{r},t)$
\begin{equation} \label{eq:fieldorthog}
\iint d\vec{r} E_{\mbox{i}}^*(\vec{r},t) E_{\mbox{e}}(\vec{r},t) = 0 \ ,
\end{equation}
as long as the anti-symmetry (\ref{eq:phisym}) of the phase modulation function holds valid. Then after reflection from the cavity, the modulated light field becomes \cite{hansch1980}
\begin{eqnarray} \label{eq:nreflection}
E_{\mbox{r}}(\vec{r},t)&=&\left\{\sqrt{R_1} e^{i\varphi(\vec{r})}
 -A\frac{T_1}{\sqrt{R_1}}\frac{e^{i\delta}\sqrt{R}}{1-e^{i\delta}\sqrt{R}}\right\} E_{\mbox{i}}(\vec{r},t) \nonumber \\
&\equiv& \left[\chi e^{i\theta} + i\sin\varphi(\vec{r})\right] \sqrt{R_1}\ E_{\mbox{i}}(\vec{r},t) \ ,
\end{eqnarray}
in which
\begin{equation}
\chi e^{i\theta}\equiv \cos\varphi(\vec{r})- A\frac{T_1}{R_1}\frac{e^{i\delta}\sqrt{R}}{1-e^{i\delta}\sqrt{R}} \ .
\end{equation}

After substituting Eqs. (\ref{eq:nreflection}) and (\ref{eq:phiexample}) into Eq. (\ref{eq:errorsignal}), one arrives at
\begin{eqnarray}\label{eq:esstrong}
\varepsilon(\delta) &=& \iint_{S_1} d\vec{r} R_1 |(\chi e^{i\theta}+i\sin\varphi_0)E_{\mbox{i}}(\vec{r},t)|^2 \nonumber \\
&& -\iint_{S_2} d\vec{r} R_1|(\chi e^{i\theta}-i\sin\varphi_0)
E_{\mbox{i}}(\vec{r},t)|^2 \nonumber \\
&=& 4R_1I_0\ \sin\varphi_0 \ \mbox{Im}\left(\cos \varphi_0- A \frac{T_1}{R_1}\frac{e^{i\delta}\sqrt{R}}{1-e^{i\delta}\sqrt{R}}\right)\ ,
\end{eqnarray}
from which it follows that the error signal for cavity locking varies as the cavity detunes from its peak resonance (Fig. \ref{fig:esstrong}) when $\sin\varphi_0\ne 0$.

From Fig. \ref{fig:esstrong} it follows that the size of the error signal can be almost twice that of the PDH one with optimized modulation depth of $\beta=1.08$, for which the PDH scheme needs to pay the price of high-order
harmonic r.f. modulation side-bands in the optical spectrum of the incident beam. We stress that the error signal in this scheme is generated as the differential photo-current from the two quadrants of the detector and,
thereby, low-frequency optical and detection noises can be substantially suppressed. Despite this, the error signal may still be contaminated by residual low-frequency noises. For cases where these low-frequency noises are
detrimental to laser stability, a possible solution could be to use a dual-frequency incident light field with two equal-strength components such that the error signal is up-shifted to r.f. bands. When both optical frequency
components are chosen to simultaneously resonate in the cavity, the r.f. error signal reads
\begin{eqnarray}\label{eq:esdualf}
\varepsilon(\delta) &=& 4R_1I_{AC}\ \sin\varphi_0 \ \chi\sin\theta  \nonumber \\
&=& 4R_1I_{AC}\ \sin\varphi_0 \ \mbox{Im}\left(\cos\varphi_0- A\frac{T_1}{R_1}\frac{e^{i\delta}\sqrt{R}}{1-e^{i\delta}\sqrt{R}}\right)\ ,
\end{eqnarray}
wherein $I_{AC} = \bar{I} \cos \Omega t$ with $\bar{I}$ being the average power of the incident beam and $\Omega$ the optical frequency span between the two frequency components. A dual-frequency light beam may be created by a
dual-frequency laser \cite{Chou2007} or by use of an acousto-optic modulator and a single-frequency laser \cite{yang2019}.

Last but not least, let discuss the differences between the studied cavity locking method and the tilt-locking one, the latter of which is modulation free \cite{Shaddock2000}. By slightly tilting or laterally displacing the
input beam of a well-aligned and mode-matched cavity, a TEM$_{10}$ mode may be generated to interfere with the fundamental TEM$_{00}$ mode to produce an error signal. This particular way to create error signal makes the system
very sensitive to light beam drifting noises as evidenced by the fact that unexpected beam drifting is indistinguishable from deliberate beam alignment for error signal generation; therefore, double pass configuration with
travelling-wave cavities must be used to suppress these noises \cite{Chabbra2021}. This leads to the major disadvantage of the tilt-locking scheme, i.e., its incompatibility with the standing-wave cavity configuration used by
the PDH method. On the contrary, spatial phase modulation in the studied method produces the error signal in a different way and hence is not so sensitive to beam drifting as the tilt locking one; in addition, residual beam
drifting noises can be further corrected as discussed in the previous text. Therefore, our proposed scheme needs not stick to travelling-wave cavities and allows one to implement laser stabilization with FP cavities, just as
the PDH scheme does. In other words, both the techniques of spatial modulation (our method) and temporal modulation (the PDH one) of optical phase front can share the same optical platform for laser stabilization, which will
save much time and man power for cavity design.

To conclude, we have studied optical cavity locking for laser stabilization through spatial modulation of the phase front of an optical carrier. A theoretical description of the underlying principle has been developed for the cavity locking method with a special attention paid to RAM noises caused by experimental imperfections. While the studied method has the common advantages of the PDH method and the tilt-locking one, it can provide a more artful way to eliminate RAM noise in phase modulation. This method can implement laser stabilization with FP cavities like the PDH one, but giving larger error signals and cleaner spectra for the carriers than the later. In situations where cost and portability are a practical issue, the studied scheme allows one to realize laser stabilization in the PDH optical platform without use of expensive and bulky devices such as signal generators and electro-optic modulators.

\textbf{Funding} This work was supported by the National Natural Science Foundation of China (12074110).



\textbf{Disclosures} The authors declare no conflicts of interest.

\textbf{Data availability} Data underlying the results presented in this paper are not publicly available at this time but may be obtained from the authors upon reasonable request.






\end{document}